%% file: main.tex
\documentclass[conference]{IEEEtran}
\IEEEoverridecommandlockouts
\usepackage{cite}
\usepackage{comment}
\usepackage{amsmath,amssymb,amsfonts}
\usepackage{algorithmic}
\usepackage{graphicx}
\usepackage{subcaption}
\usepackage{float}
\usepackage{textcomp}
\usepackage{xcolor}
\def\BibTeX{{\rm B\kern-.05em{\sc i\kern-.025em b}\kern-.08em
    T\kern-.1667em\lower.7ex\hbox{E}\kern-.125emX}}

\newcommand{\mycomment}[1]{}

\begin{document}

\title{Efficient and Portable Support for Overdecomposition on Distributed Memory GPGPU Platforms}

\author{
Aditya Bhosale\IEEEauthorrefmark{1},
Anant Jain\IEEEauthorrefmark{2},
Shourya Goel\IEEEauthorrefmark{2},
Ritvik Rao\IEEEauthorrefmark{1},
Peddoju Sateesh Kumar\IEEEauthorrefmark{2},
Laxmikant Kale\IEEEauthorrefmark{1}\\

\IEEEauthorblockA{\IEEEauthorrefmark{1}
University of Illinois Urbana-Champaign, Urbana, IL, USA\\
Emails: \{adityapb, rsrao2, kale\}@illinois.edu
}

\IEEEauthorblockA{\IEEEauthorrefmark{2}
Indian Institute of Technology, Roorkee, Roorkee, India\\
Emails: \{anant\_j, shourya\_g, sateesh\}@cs.iitr.ac.in
}

}

\maketitle



\begin{abstract}

Overdecomposition has emerged as a powerful and sometimes essential technique in parallel programming. Many application domains or frameworks, including those based on adaptive mesh refinements, or tree codes use it. Charm++ is a parallel programming system which has demonstrated the utility of overdecomposition for many applications and in multiple contexts. However, the emergence of GPGPUs as a dominant compute component has created some real and perceived challenges for this paradigm, especially regarding the higher overhead brought about by overpartitioning --- having multiple objects assigned to the same GPGPU device. We address this issue as well as the issue of portability by developing techniques and software that demonstrate that overdecomposition can be efficiently and productively supported on combinations of GPU vendor types, and interconnection networks. 
\end{abstract}


\input{sections/introduction}
\input{sections/background}
\input{sections/related_work}
\input{sections/execution_model}
\input{sections/comm_layer}
\input{sections/results}
\input{sections/conclusion}

\bibliographystyle{plain}
\bibliography{group, references}

\end{document}

%% file: sections/introduction.tex
\section{Introduction}

Modern HPC has entered an era of unprecedented architectural diversity. The leading systems on the TOP500 list are distributed across at least three GPU vendors: Frontier and El Capitan are powered by AMD accelerators, Aurora by Intel, and Jupiter Booster, Eagle, and Alps by NVIDIA GPUs. Each vendor ships its own native programming model, HIP, SYCL, and CUDA respectively, and each uses a distinct network stack. For application developers and runtime system designers, this diversity has turned portability from a convenience to a first-class requirement.

Real applications increasingly exhibit dynamic behavior, for example, adaptive mesh refinement, multi-physics simulations, and irregular algorithms whose loads are data dependent and hard to statically predict. Hardware contributes another source of imbalance, from performance variability between identical GPUs, to thermal throttling, and network contention. The bulk-synchronous model that dominates large-scale GPU codes offers no inherent mechanism for tolerating this dynamism. Asynchronous Many Task (AMT) runtimes such as Charm++~\cite{CharmppOOPSLA93, sc14charm}, HPX~\cite{HPX}, Legion~\cite{Legion}, etc. address this gap by decoupling the application's expression of parallelism from its hardware mapping.   

Charm++ addresses adaptivity through overdecomposition, ie. dividing the computation into many more migratable objects (chares) than processing elements (PEs), and letting the runtime schedule, overlap, and migrate them with minimal application intervention. These benefits are well-established on CPU-based systems and have been validated at scale in production codes such as NAMD~\cite{NamdJCC05} and ChaNGa~\cite{2007_ChaNGaScaling}. However, extending the same model to GPUs has a fundamental challenge: GPUs achieve peak efficiency with large, high-occupancy kernels and bulk data transfers, while overdecomposition deliberately partitions work into many smaller chunks. The concern is that the resulting fine granularity will manifest as reduced occupancy, increased kernel launch overheads, and synchronization costs that will together outweigh the benefits of overdecomposition. 


Previous work from Choi et al.~\cite{Choi22} quantified these costs for GPU-aware Charm++ applications on NVIDIA Summit, using a single-PE-per-GPU configuration, and proposed application-level mitigations such as kernel fusion to coalesce fine-grained packing/unpacking kernels with the computation kernel, and CUDA Graphs to amortize kernel launch overhead for iterative applications. While these techniques are effective, they require the application developer to restructure code around the runtime system's granularity. Their study was also limited to a single GPU vendor and a single communication stack. Whether overdecomposed GPU execution can be supported portably from a single source across multiple GPU vendors and communication stacks, whether its overhead can be managed transparently at the runtime layer instead of the application layer, and how it scales with the overdecomposition factor on modern accelerators remain open questions.

In this paper we take a step towards answering them. We extend Charm++ with a portable GPU communication layer built on the Lightweight Communication Interface (LCI)~\cite{LCI} that operates over both ibverbs and libfabric, paired with Kokkos~\cite{KokkosCore2014} for execution portability. The result is a single-source Charm++\-/Kokkos environment in which an overdecomposed application runs unchanged on NVIDIA and AMD GPUs.

Within this environment we develop two application agnostic improvements aimed at managing the cost of fine-grained execution on GPUs. The first uses non-blocking GPU streams so that each chare's fine-grained packing/unpacking kernels overlap with the computation kernels of other chares mapped to the same device. The second generalizes the runtime's execution model into a hierarchy in which multiple processes are mapped to each GPU and multiple worker threads (PEs) are mapped to each process. Mapping multiple PEs to a shared GPU allows kernels to be submitted from these PEs in parallel, mitigating the launch side serialization that a single-PE design incurs as the overdecomposition factor grows. Both techniques are implemented entirely in the runtime requiring no application level restructuring, and complement the application-level techniques explored in prior work. On top of this foundation, we conduct a detailed empirical study of overdecomposition cost, measuring how the application overhead scales with the overdecomposition factor across vendor platforms, and characterizing the trade-offs across execution model configurations.

This paper makes the following contributions:
\begin{enumerate}
    \item A portable GPU runtime for Charm++ built on LCI and Kokkos, supporting NVIDIA and AMD GPUs over both libverbs and libfabric from a single source.
    \item Two application-agnostic, runtime-level techniques for managing the cost of overdecomposition on GPUs: non-blocking streams that overlap each chare's pack/unpack with other chares' computation, and a hierarchical execution model with multiple processes per GPU and multiple worker threads per process that enables parallel kernel submission to a shared device.
    \item A detailed empirical study of overdecomposition cost on AMD and NVIDIA GPUs, attributing per-chare overhead to its sources, characterizing how it scales with the overdecomposition factor, and analyzing the trade-offs across execution model configurations.
\end{enumerate}

%% file: sections/background.tex

%% file: sections/related_work.tex
\section{Background and Related Work}

\subsection{Performance portability frameworks}

Kokkos~\cite{KokkosCore2014}, RAJA~\cite{raja}, and SYCL provide single-source abstractions that compile to vendor-native backends, allowing parallel kernels to target CPUs and GPUs from any of the major vendors without source modification. These frameworks have been adopted at scale across the DOE exascale machines. By design, they scope themselves to within-node execution and do not address inter-node communication or runtime adaptivity. Our work uses Kokkos for execution portability and complements it with a portable communication layer.

\subsection{Asynchronous many-task runtimes on GPUs}

Several AMT runtimes have been extended to multi-vendor GPU execution. Uintah~\cite{uintah} has been ported to AMD, NVIDIA, and Intel GPUs through Kokkos backends, demonstrating single-source portability for an asynchronous many-task framework across all three GPU vendors.

\subsection{Charm++ and Overdecomposition}

The specific AMT we extend here is Charm++, which pioneered the idea of overdecomposition.  In charm++, programmer decomposes the work and data into C++ objects called chares, that are organized into one or more multi-dimensional collections. Chares are assigned to processors by the Charm++ runtime. The application developed does not need to be aware of the location (processor) where a chare lives --- A chare communicates with another chare using its collection name and index, sending an asynchronous method invocation towards it. So, a call such as A[i,j].foo(params) immediately returns, while a message containing the params data is send towards the last known location where the (i,j) member of chare collection A lives by the system, where the foo method is eventually scheduled by a user-space queue based scheduler. 

The ability to migrate chares across processors confers the ability to dynamically balance loads to an introspection (instrumentation) based runtime system. 

\mycomment{Need a paragraph on applications developed using Charm++ as well runtime optimizations, as evidence of utility of overdecomposition}


%% file: sections/execution_model.tex
\section{Execution Model}



\begin{figure}
    \centering
    \includegraphics[width=.9\linewidth]{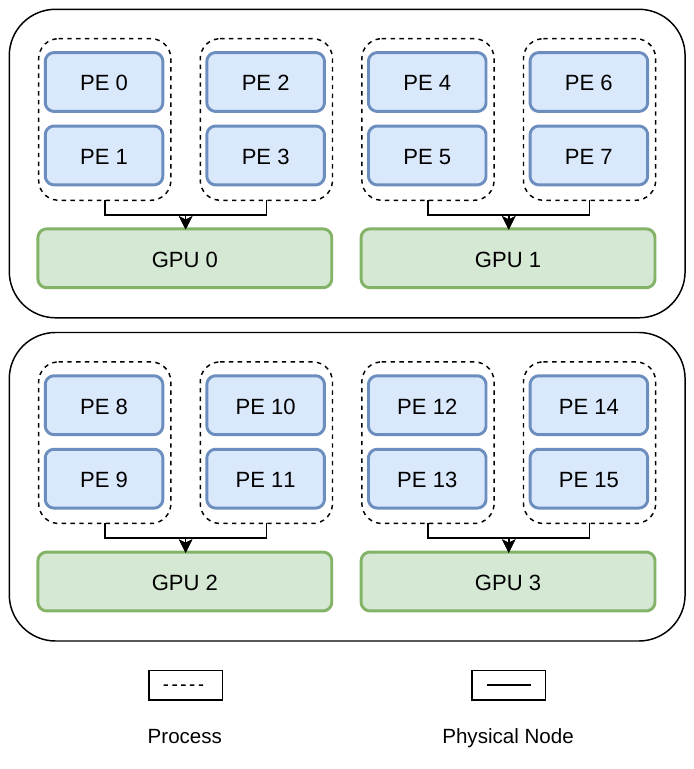}
    \caption{Execution model for GPU-aware Charm++ applications}
    \label{fig:execution-model}
\end{figure}

We organize the runtime into a hierarchy of user-level threads, processes, and GPUs, illustrated in Figure~\ref{fig:execution-model}. Each GPU is shared by one or more processes, and each process contains one or more worker threads. Chares are scheduled onto PEs by the Charm++ runtime, and each chare issues its GPU kernels to the GPU associated with its process. PEs within the same process share a CUDA or HIP driver context, enabling direct device-to-device copies between chares. Cross-process GPU messages uses CUDA or HIP IPC on the same node and GPU-aware RDMA across nodes.

\subsection{Managing Overdecomposition} 

\mycomment{ Section message/outline: (having many objects per pe or gpu creates challenges. Use GPU device as unit/range of allocation. How significant are kernel startup overheads. How to overcome those? using multiple PEs and chares helps.
Description of how kokkos was woven in..
Challenges. (synchrony/asynchrony..)
}

The idea of overdecomposition means there will be many objects (Chares), each with its own kernels running on each GPU and each PE. How many should there be? In the non-GPU context, Charm++ typically needs about 8-16 chares per PE (a PE is roughly the same as a core. It is a core that has a charm scheduler running on it). This allows the load balancer adequate flexibility in moving some number of chares into or away from a core to restore load balance. Any smaller than that, and one suffers the effects of quantization. (An analogy is distributing rocks to buckets, to keep all the buckets of equal weight. Heavy chunks would hinder the ability to balance buckets). 

An important question arises when we consider GPUs: should a core still be the right unit of scheduling? Since there are hundred+ cores on modern nodes, it will be an extremely fine-grained computation if we require 8-16 chares per PE. An important insight is that computing load in modern GPU based applications is carried mostly on GPUs, and not on CPU cores. The cores on the host are used more as a coordination unit and to do small unavoidable computations. Thus, we conclude that overdecomposition should be done with respect to GPUs -- putting 8-16 chares per GPU device. Our objective, for this paper, then becomes an easier one: can we sustain an adequately fine grainsize to generate 8-16 chares per GPU without undue overhead? What are the obstacles, and what techniques can be developed to overcome them. 

\mycomment{cost of overdecomposition section text and experiments moves here.}











It has been known that a single kernel latency on modern GPUs is large -- of the order of 10-20 microseconds. That may seem like  a short time, but especially for overdecomposed kernel, that overhead may be prohibitive. 

\mycomment{ To validate this data point, and to provide a reference point for our other experiments, we ran a simple experiment in which a CPU core, i.e. a PE in charm++ terminology, fires an empty kernel, waits for its completion, and then fires the next kernel, repeatedly. We see that the kernel startup and completion latency is about 10 microseconds. BTW: this experiment can be done with a blocking CUDA call, or with  HAPI callback that you wait on.. the two might provide differential useful data points) [Keeping this in comments until we do this experiment, on any one of the gpus available to us]}

However, we hypothesize that this latency is composed of parts that can be overlapped in various ways. In particular, there may be latency associated with the work done on the host core, and separately in device runtime which is scheduling and initializing each kernel. 

To assess the potential of such overlap, we construct a microbenchmark to study the effect of overdecomposition on end-to-end kernel runtimes. Here, the work is divided into a variable number of kernels. To isolate GPU runtime overhead, all kernels are enqueued into their respective streams up front and then released simultaneously by having each stream wait on a host-mapped synchronization flag, using \texttt{cuStreamWaitValue32}.\mycomment{releasing them using what call? mentioning it will be good for completeness}

\begin{figure}
    \centering
    \includegraphics[width=1\linewidth]{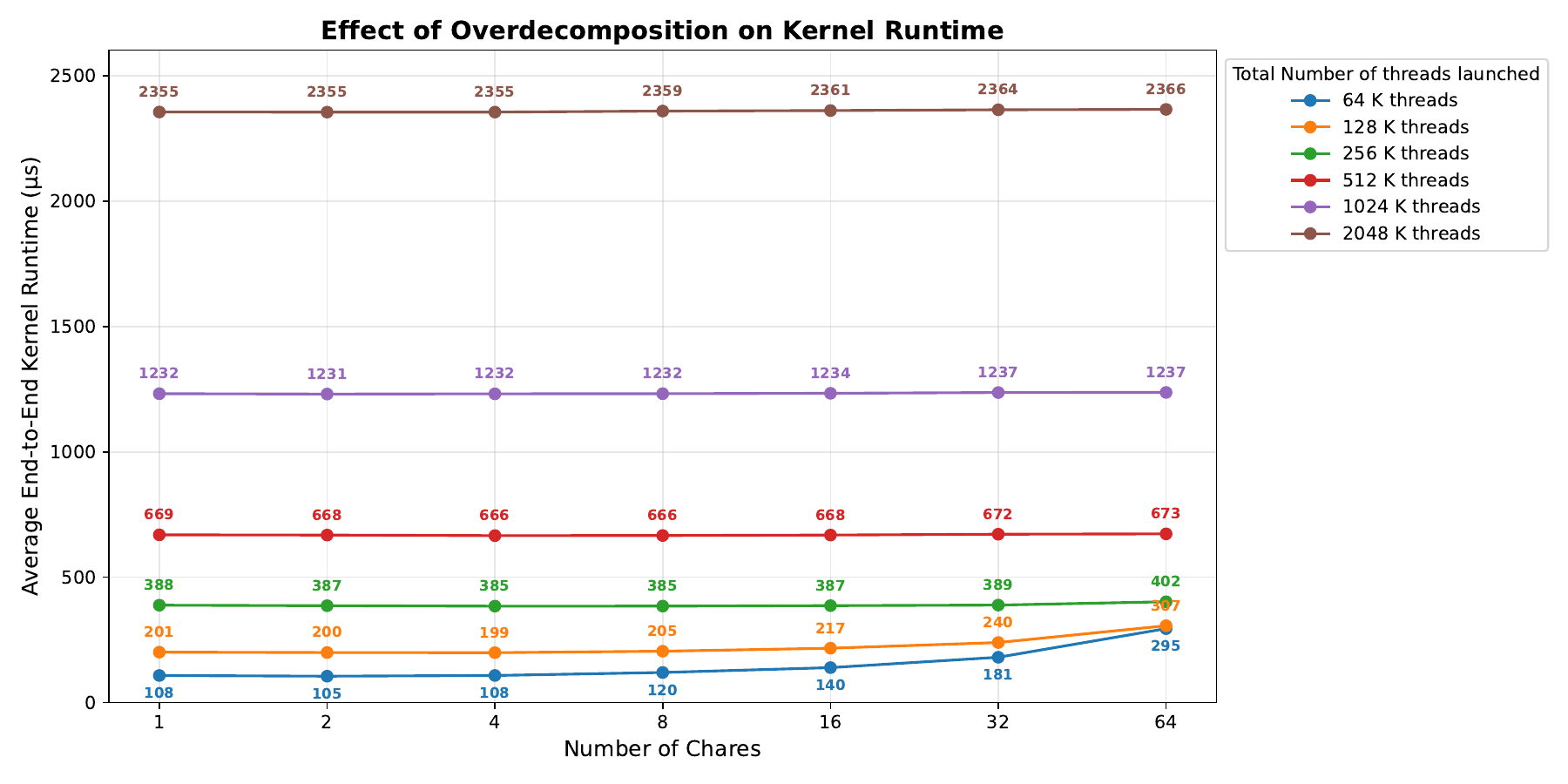}
    \caption{Effect of overdecomposition on Kernel runtimes}
    \label{fig:fullOverlap}
\end{figure}

The resultant data is compiled in Figure \ref{fig:fullOverlap} with the total number of threads meaning the number of CUDA threads in CUDA kernel launch. The Results show that overdecomposition, in this highly controlled situation has pretty low overhead. For the workload size as low as \(256 K\) threads , the completion times remain broadly similar across configurations, ranging from \(388~\mu s\) when the work is launched from 1 chare to \(402~\mu s\) when the work is launched from 64 chares, indicating that overdecomposition introduces relatively little overhead in this controlled setting. For example, the smallest configuration completes in about \(108~\mu s\), increasing only to about \(120~\mu s\) when the same work is split across 8 kernels, and to \(296~\mu s\) at 64 kernels. The larger increase at the highest ODFs is likely due to CUDA runtime overheads, including scheduler pressure and limits on kernel concurrency.

Submitting kernels ahead of time to streams, and only releasing them on signal, is a synthetic mechanism to demonstrate the potential. In reality, an asynchronous task-based runtime will be submitting kernels at arbitrary times based on availability of data.  Further, if there is a component latency borne by the CPU core, it may be better to use multiple CPUs for each GPU. After all, many more cores exist on a node than GPUs. To study this aspect, we ran a new benchmark.

\begin{figure}
    \centering
    \includegraphics[width=1\linewidth]{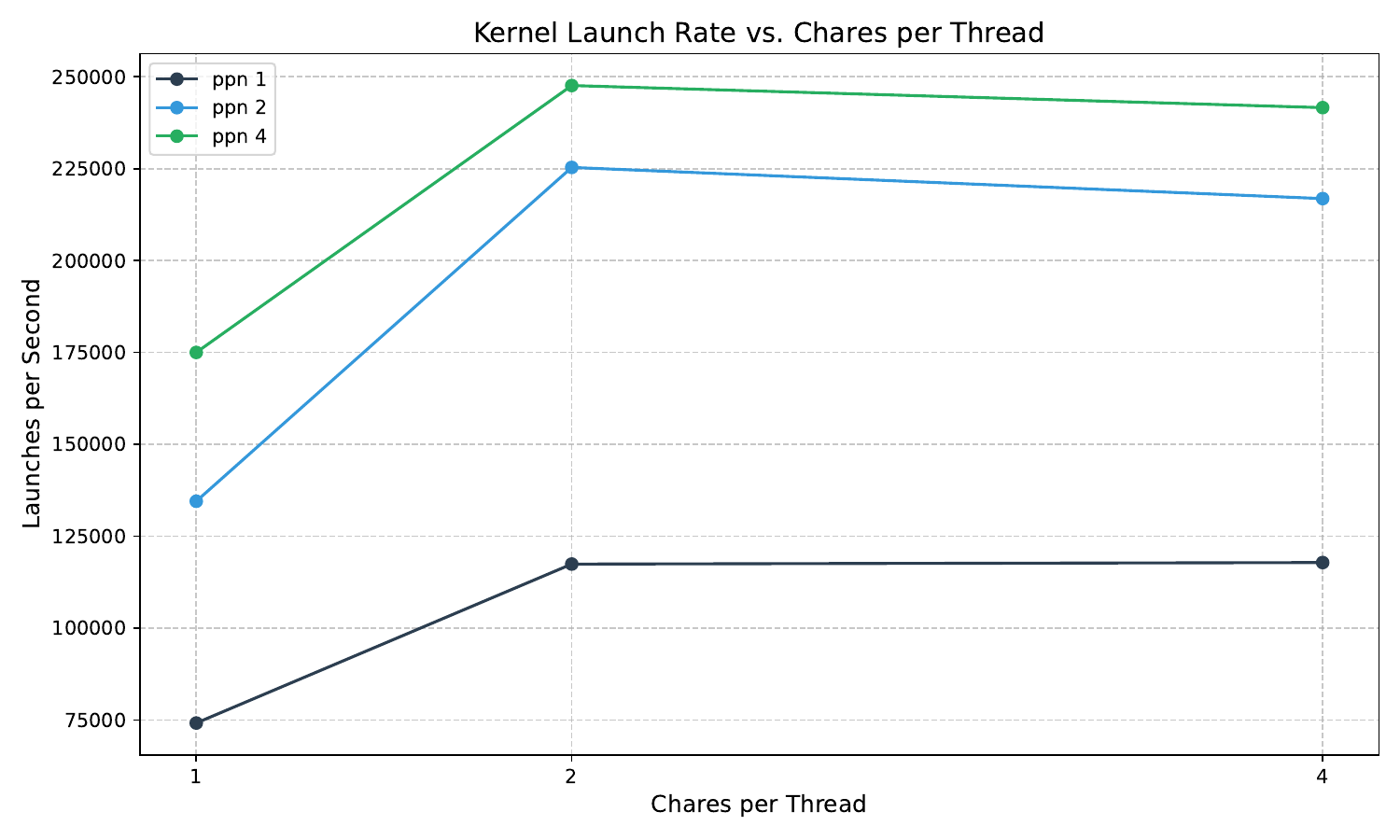}
    \caption{Rate of Kokkos kernel launches on A40 vs. number of chares per thread}
    \label{fig:kernel_launch_rate}
\end{figure}

This benchmark involves rapidly firing kernels from a single process to a single GPU, while varying the number of threads and the number of chares per thread. The benchmark shows how overdecomposition and multithreading can increase the use of a GPU for fine-grained kernels. In our benchmark, each kernel is a Kokkos kernel that occupies all the SMs/compute units of a device, but otherwise does not actually do any work. When a kernel completes, we use a stream (1 per chare) to trigger a HAPI callback that will fire the next kernel. Fig. \ref{fig:kernel_launch_rate} shows the number of kernels launched per second with 1-4 chares per PE and 1-4 PEs. From this experiment, there is a clear advantage to using 2 chares per thread as opposed to 1, as the Charm++ scheduler can switch to the next chare and fire a kernel while the first chare waits on a callback. However, using 4 or more chares per thread shows no advantage. This may be because of the overhead of the Charm++ scheduler (running on the host CPUs) is large enough that the first kernel launch will likely have returned by the time a 4th chare's kernel is ready to launch. \mycomment{We were unable to run the benchmark with more than 4 chares per thread or 4 threads per process (ppn) because of an error caused by multiple HAPI callbacks being placed into the same scheduler queue. We plan to improve the queue structure in Reconverse to solve this issue.}


\subsection{Using Kokkos in charm++ runtime}
Kokkos is primarily designed around a synchronous execution model such as MPI, and integrating it effectively with Charm++ requires additional considerations to preserve asynchronous execution and overlap. One source of overhead arises from implicit synchronization in operations such as temporary buffer deallocation and APIs like \texttt{Kokkos::resize}, which internally invoke \texttt{Kokkos::fence} because of Kokkos's reference-counting-based memory management. In a multithreaded Charm++ execution model, these implicit fences can unnecessarily stall unrelated chares. To avoid such synchronization points, we make frequently used \texttt{Kokkos::View}s persistent members of the class implementation and obtain scratch space from preallocated buffers instead of dynamically allocating temporary buffers during execution.

Another challenge stems from Kokkos's functor caching behavior, which can introduce additional synchronization across kernel launches. Since our applications do not launch identical kernels in a tightly consecutive manner, the caching mechanism provides almost no benefit while increasing synchronization. We mitigate this behavior using Kokkos's lightweight hints.

To support concurrent execution across multiple streams, we use separate Kokkos execution spaces. In doing so, several considerations become important. For example, \texttt{Kokkos::deep\_copy} operations issued without an explicit execution space can introduce host-side synchronization, Kernels issues without an execution Space run on the default stream, which synchronizes implicitly with the rest of the streams. Additionally, initializing an execution space itself is relatively expensive because it involves device initialization and associated memory copies from host to device. To minimize this overhead, execution spaces are created once during initialization and reused throughout the lifetime of the application.


\subsection{Performance trade-offs}

Wider processes, ie. more PEs per process and fewer processes per GPU, keeps more message traffic within a single driver context, where it benefits from direct device-to-device copies. Smaller processes routes more traffic through the IPC mechanism which requires two intermediate copies into pre-allocated buffers on both the sender and receiver. Wider processes, however, incur an overhead due to thread contention on the host for kernel launches.

On NVIDIA hardware, multi-process configurations require the Multi-Process Service (MPS) for concurrent kernel execution across processes. Without MPS, kernels from different contexts time-slice. ROCm provides no equivalent service, so multi-process configurations on AMD hardware time-slice rather than execute concurrently. The benefits of multi-process-per-GPU configurations are therefore vendor-asymmetric, while the costs are not.

\subsection{Pipelined communication}
We evaluate GPU-direct communication and computation–communication overlap on 2 A40 GPUs connected at ~21 GB/s peak one sided bandwidth. 
The benchmark transfers a fixed total data volume split across a varying number of sender receiver pairs. ODF is the number of sender-receivers pairs.

To reflect realistic application behavior we append a small kernel launch at the end of each receive. The total end-to-end time thus captures both increase in transfer time because of overdecomposition and benefit due computation communication overlap.

\begin{figure}
    \centering
    \includegraphics[width=1\linewidth]{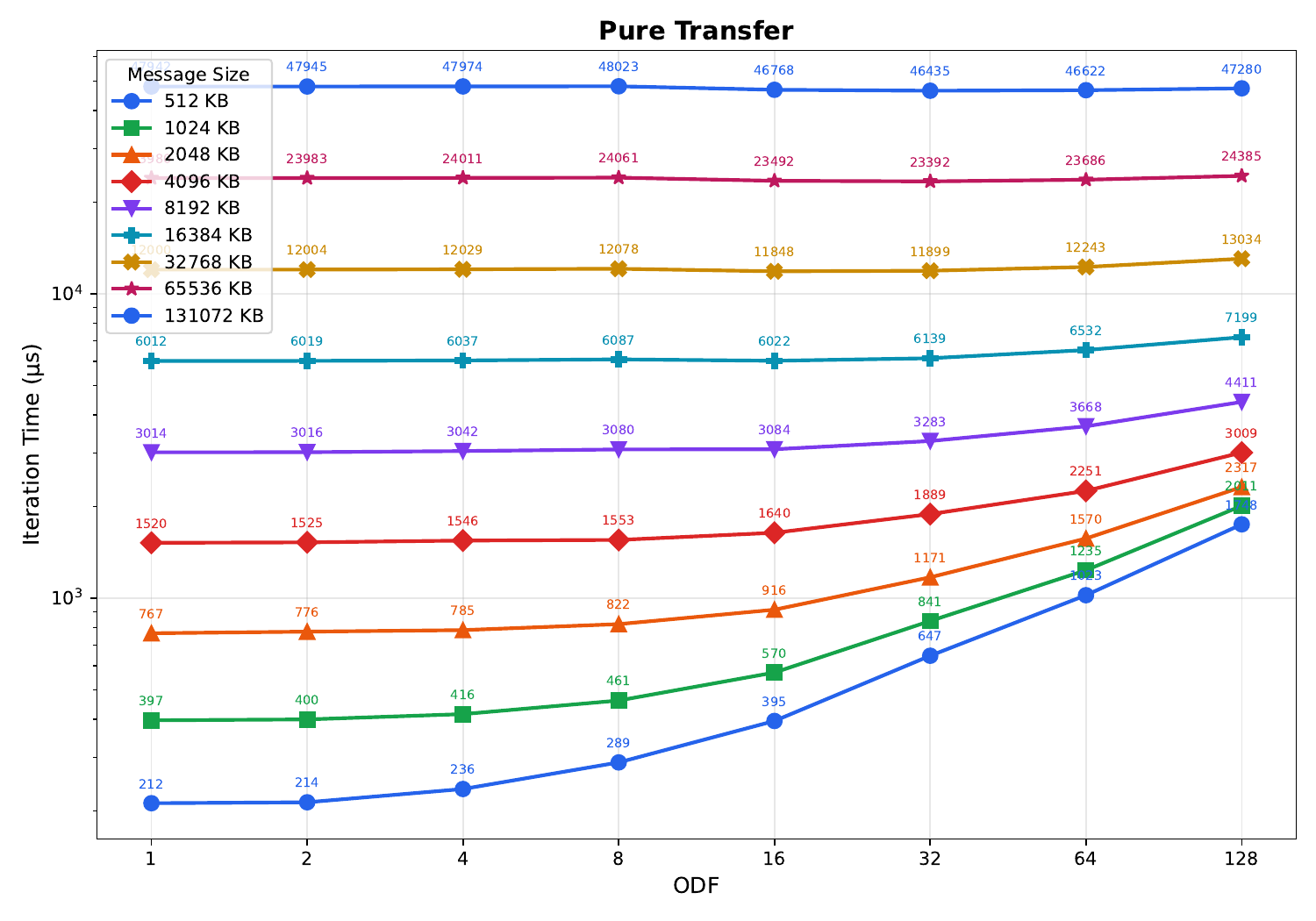}
    \caption{Effect of overdecomposition on communication}
    \label{fig:overdecomposed-comms}
\end{figure}

Figure~\ref{fig:overdecomposed-comms} shows for the data sizes for a typical application, the communication only starts to climb up when the amount of overdecomposition goes beyond 16, which is enough for load balancing to work effectively.

\begin{figure}
    \centering
    \includegraphics[width=\linewidth]{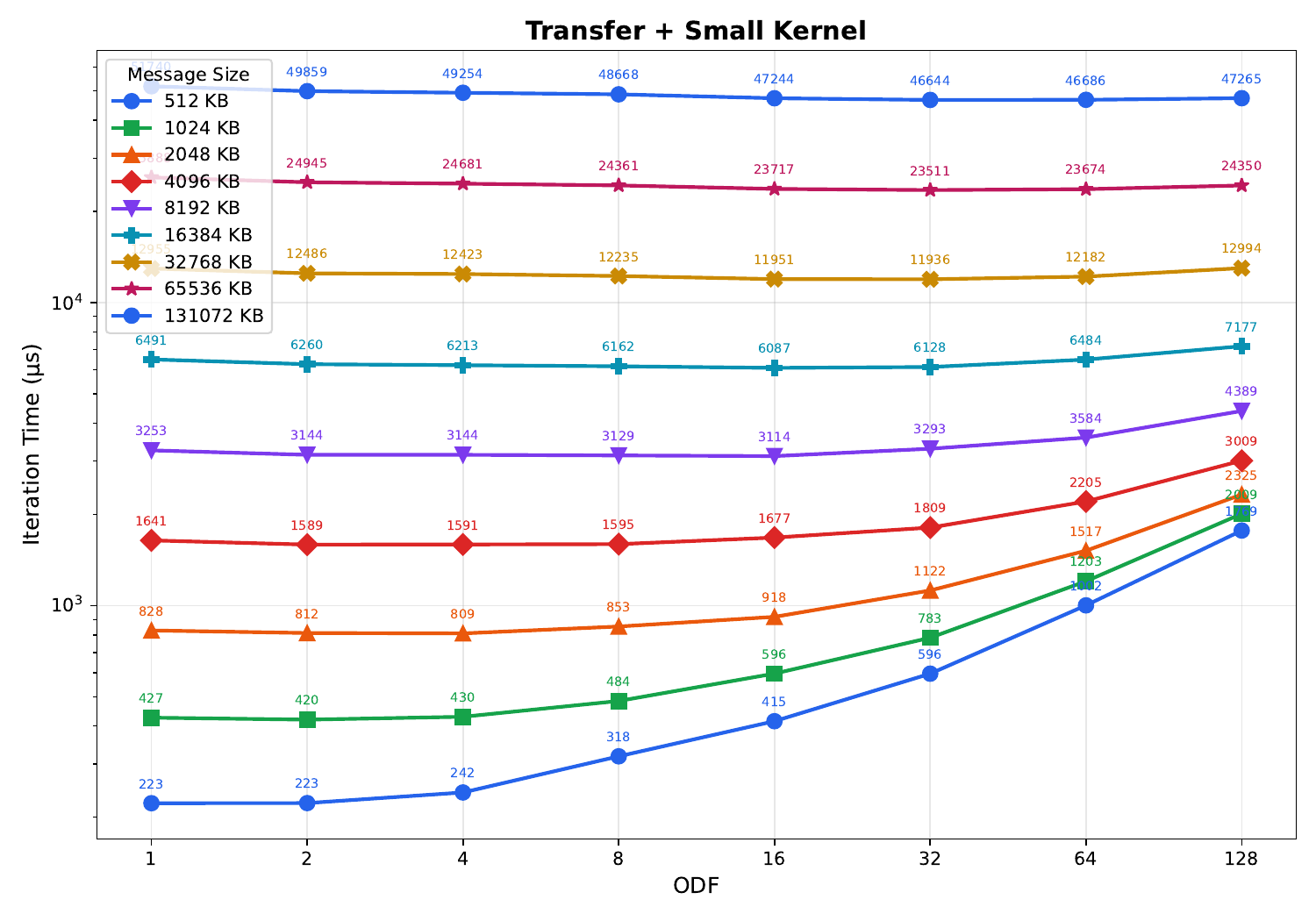}
    \caption{Effect of Communication-Computation Overlap}
    \label{fig:overdecomposed-comms-compute}
\end{figure}

Figure \ref{fig:overdecomposed-comms-compute} shows the effect of adding an O(n) computation kernel at the end of the transfer. Total walltime for most sizes beyond 2048 KB are reduced for some higher degree of overdecomposition depending on the data size. 

%% file: sections/comm_layer.tex
\section{Portable Communication Layer}

When one overdecomposes the work and data units, naturally the amount of data communicated is fragmented into smaller but more numerous messages. For example, in a 2D stencil computation, If you over-decompose by a factor of 4, each message is of half the size as before. To complicate mattes further, some of these messages may go to chares on the same device, some may go to a neighboring device on the same physical node and some may go to another node. To ensure overdecomposition overheads are reduced, all three paths must be individually optimized. The charm++ runtime can identify the destination chare's location --- this requires a runtime option that pre-fills the location table. So it is possible to know which code path to use and send it via the best protocol. 

\subsection{Transport selection}

\begin{figure}
    \centering
 
    \begin{subfigure}[t]{\linewidth}
        \centering
        \includegraphics[width=\linewidth]{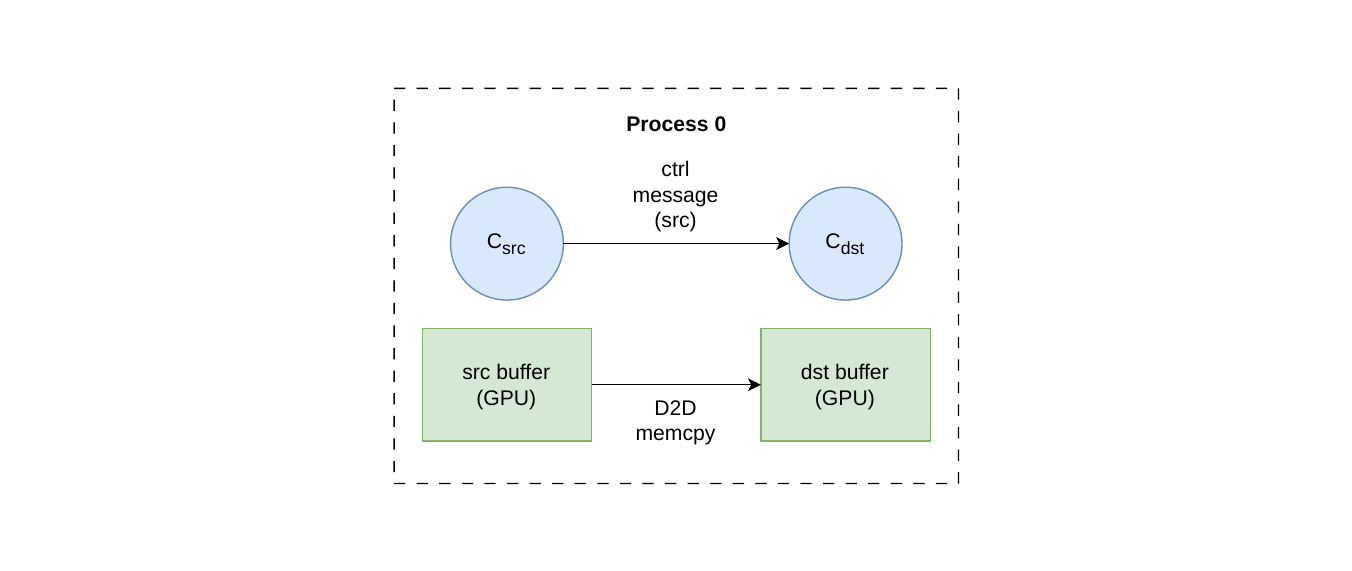}
        \caption{Intra-process GPU communication}
        \label{fig:gpu-comm:intra-process}
    \end{subfigure}
 
    \vspace{1em}
 
    \begin{subfigure}[t]{\linewidth}
        \centering
        \includegraphics[width=\linewidth]{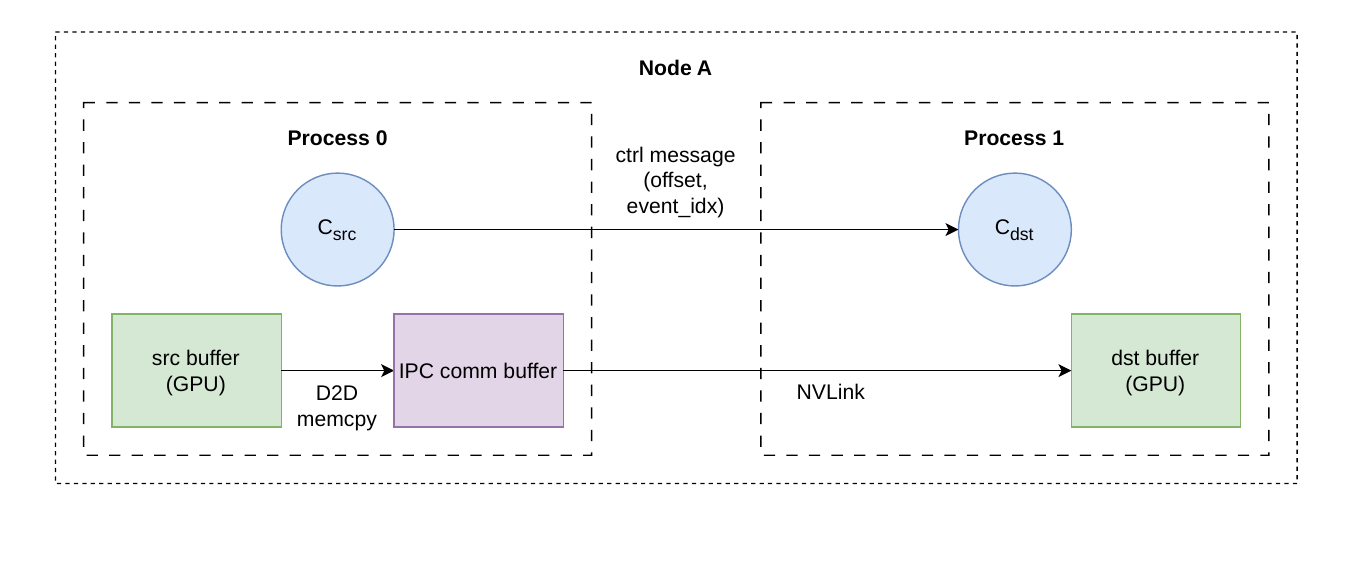}
        \caption{Intra-node, inter-process GPU communication}
        \label{fig:gpu-comm:intra-node}
    \end{subfigure}
 
    \vspace{1em}
 
    \begin{subfigure}[t]{\linewidth}
        \centering
        \includegraphics[width=\linewidth]{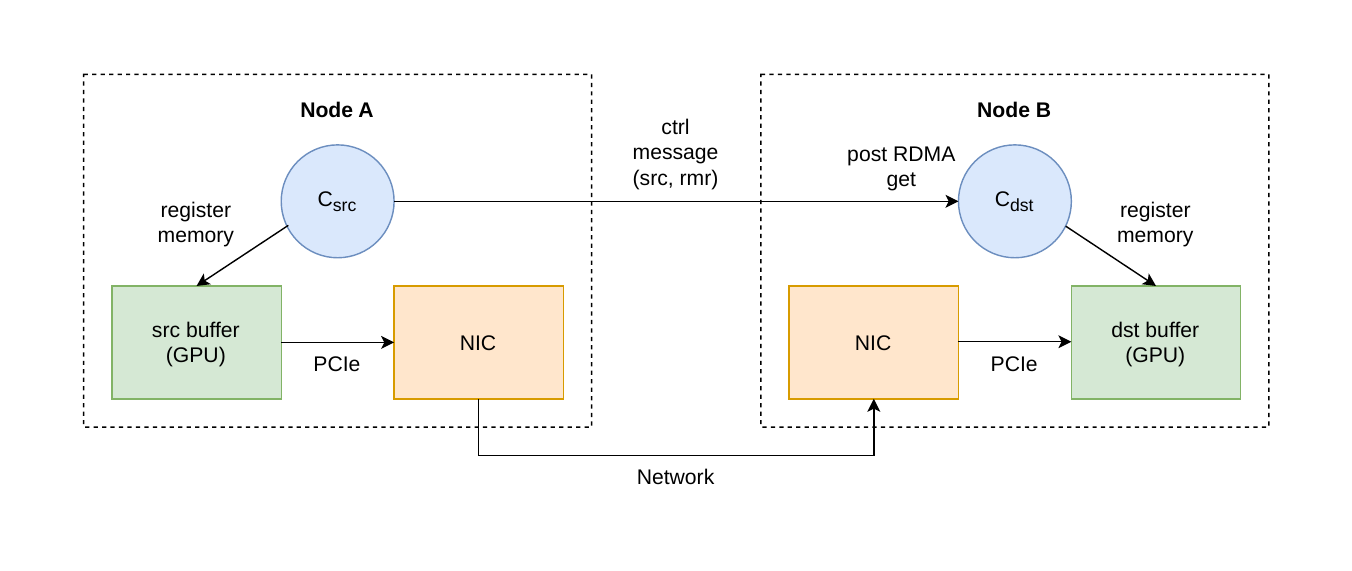}
        \caption{Inter-node GPU communication}
        \label{fig:gpu-comm:inter-node}
    \end{subfigure}
 
    \caption{GPU communication protocols in Charm++}
    \label{fig:gpu-comm}
\end{figure}

When a chare sends a device zerocopy message, the runtime selects one of three transport paths based on the location of the source and destination chares. The three transport paths are illustrated in Figure~\ref{fig:gpu-comm}.

\begin{enumerate}
    \item \textbf{Intra-process transfers} -- When source and destination chares reside on the same process, the transfer is done by a single asynchronous device-to-device copy on the stream specified in the post entry method. We use the native CUDA/HIP APIs for async memory copy to implement this. 
    \mycomment{ Since we have only one gpu in each process, this is withing the gpu memcopy, right. Explain what is used, what if anything , we needed to do to optimize this mode. Same for the other 2 modes}

    \item \textbf{Intra-node, inter-process transfers} -- When the source and destination chares are in different processes on the same physical node, the transfer is done using IPC. To avoid the high cost of opening an IPC memory handle for every message, we stage the IPC messages through a communication buffer. Each process pre-allocates a shared device communication buffer and a pool of cross process synchronization events at startup. At startup, every process also maps every peer's communication buffer and events into its own address space using POSIX shared memory to exchange handles. The transfer then involves two device-to-device copies - source buffer to the source's communication buffer, and source's communication buffer to the destination buffer. An IPC event is used on the destination process to wait for the source copy to complete before the destination copy. The chare-level control message contains the offset into the source communication buffer and the index of event pool used for synchronization.

    \item \textbf{Inter-node transfers} -- For source and destination chares on different nodes, we use one-sided RDMA. The source PE registers the source buffer and sends a control message to the destination PE with the source buffer address. The destination PE then issues a one-sided get operation that pulls the data from the source GPU into its own.
\end{enumerate}

%% file: sections/results.tex
\section{Experimental Results}
We evaluate weak and strong scaling performance of overdecomposition across three mini-applications—Jacobi2D, MiniMD, and LULESH. We compare our Charm++/Kokkos implementations of these mini-applications against MPI/Kokkos implementations across a range of ODF configurations. The nvidia runs are done on NCSA delta's A40 GPU compute nodes. Each compute node had 4 nvidia A40 GPUs, connected via PCIe Gen4. Different nodes are connected via HPE/Cray’s Slingshot 11 interconnects. The AMD runs are carried out on frontier. Each compute node on frontier has 4 AMD Instinct MI250X GPUs; each MI250X contains 2 Graphics Compute Die(GCD), so Frontier exposes 8 GCDs per node. The system uses HPE Slingshot networking interfaces.

\subsubsection{jacobi2d}
Jacobi2d is a simple stencil based application that applies the Jacobi iterative method on a 2D grid. The grid is divided among charm++ chares/MPI ranks, which communicate by halo exchanges. The application is run for 100 iterations without convergence checks.

\begin{figure}
    \centering
    \includegraphics[width=1\linewidth]{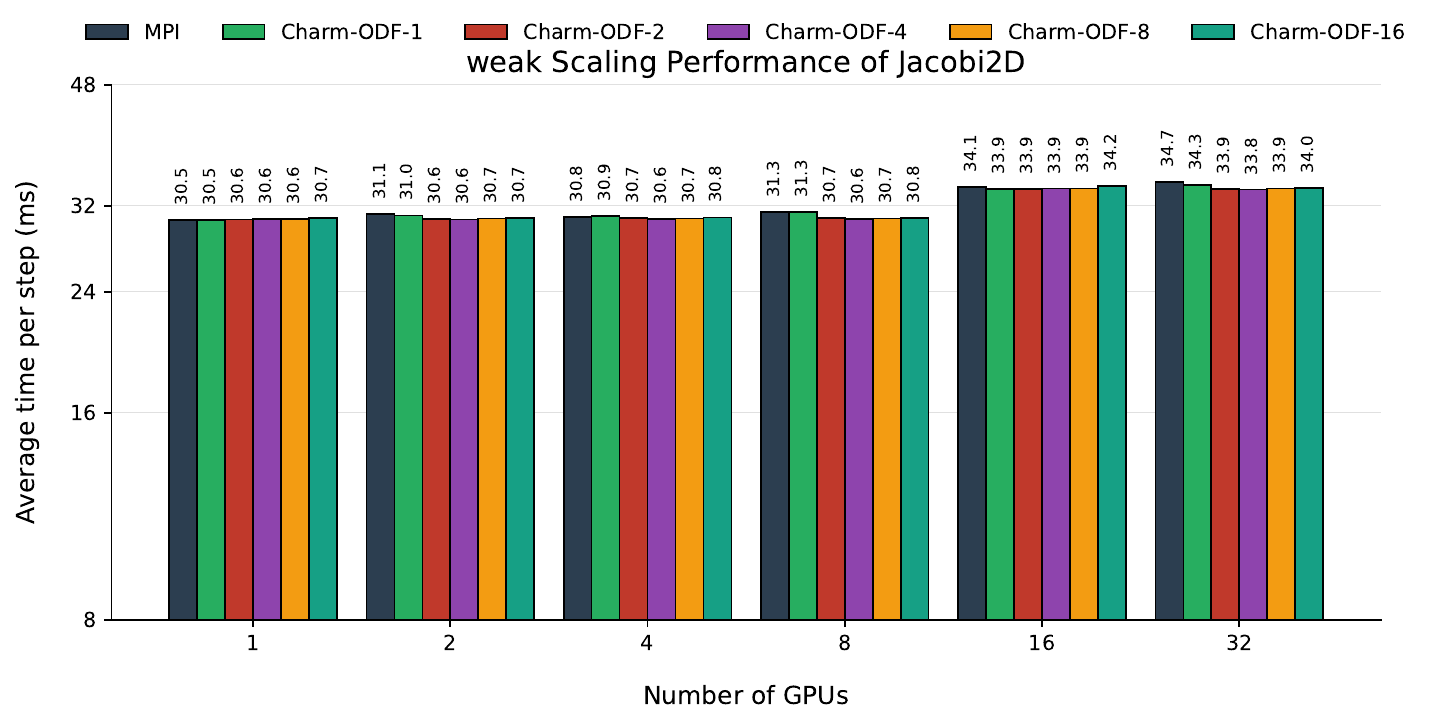}
    \caption{jacobi2d weak scaling on A40}
    \label{fig:jacobi2d_weak_scaling}
\end{figure}

\textbf{Weak scaling on Nvidia.}
Figure~\ref{fig:jacobi2d_weak_scaling} presents weak scaling results for Jacobi2D using a base domain size of \(32768 \times 32768\). As the number of GPUs doubles, the grid dimensions are alternately increased to maintain a constant workload per GPU. The experiments scale up to 32 GPUs across 8 nodes. Across the evaluated range, the different ODF configurations closely match the MPI implementation, indicating that overdecomposition introduces little additional overhead while still enabling features such as dynamic load balancing.

\begin{figure}
    \centering
    \includegraphics[width=1\linewidth]{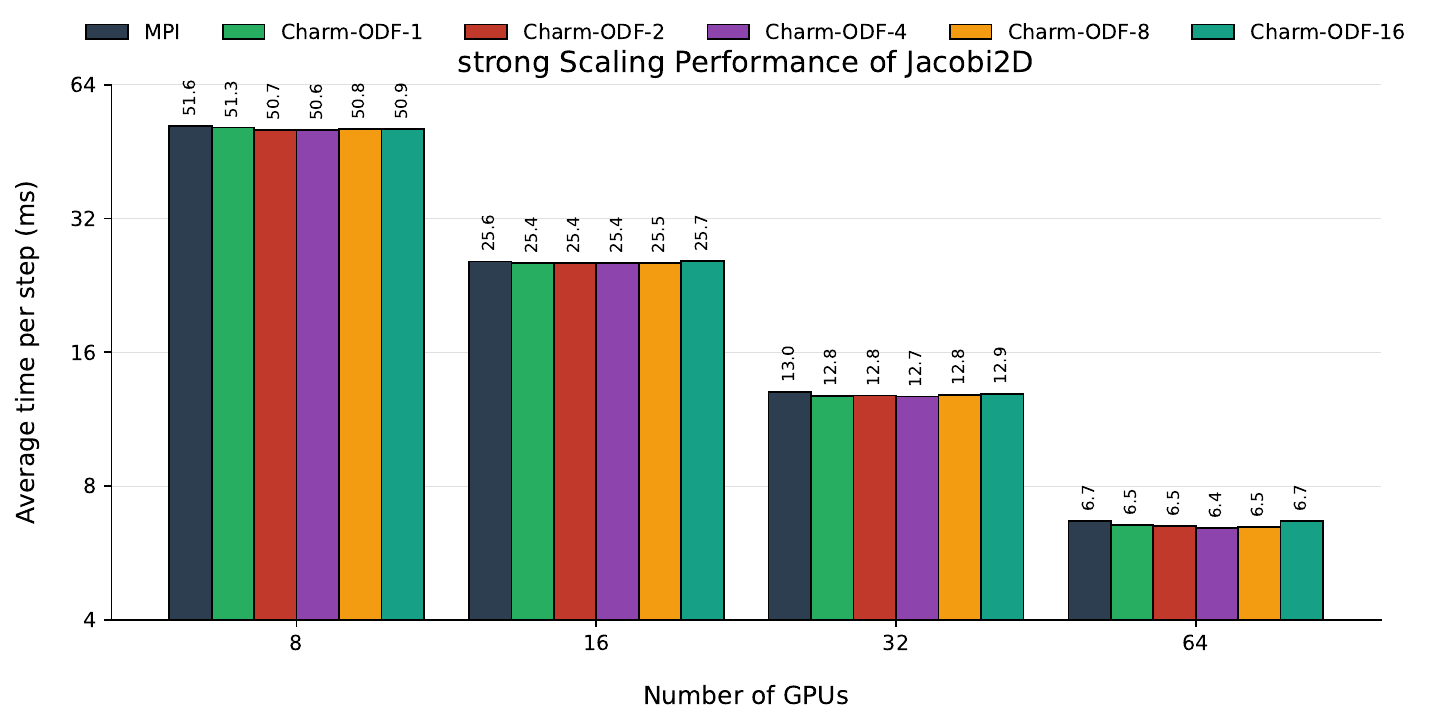}
    \caption{jacobi2d strong scaling on A40}
    \label{fig:jacobi2d_strong_scaling}
\end{figure}

\textbf{Strong scaling on Nvidia.}
For strong scaling experiments, we use a fixed domain size of \(131072 \times 98304\) and scale from 8 GPUs (2 nodes) to 64 GPUs (16 nodes). As shown in Figure~\ref{fig:jacobi2d_strong_scaling}, the ODF configurations continue to track MPI performance closely even at larger scales, consistent with the weak scaling observations.

\textbf{Weak scaling on AMD.}
On AMD systems, weak scaling is evaluated using the same base domain size as the Nvidia experiments, scaling from 1 GPU to 32 GPUs across 4 nodes. The weak scaling experiments show that time per step only has a small increase even as the number of GPUs increases, just like with the A40 runs. However, using overdecomposition shows clear benefits in execution time compared to running MPI or Charm without overdecomposition. 

\textbf{Strong scaling on AMD.}
Strong scaling on AMD uses the same fixed domain sizes as the Nvidia experiments and scales from 8 GPUs(1 node) up to 32 GPUs(4 nodes). 

\begin{figure}
    \centering
    \includegraphics[width=1\linewidth]{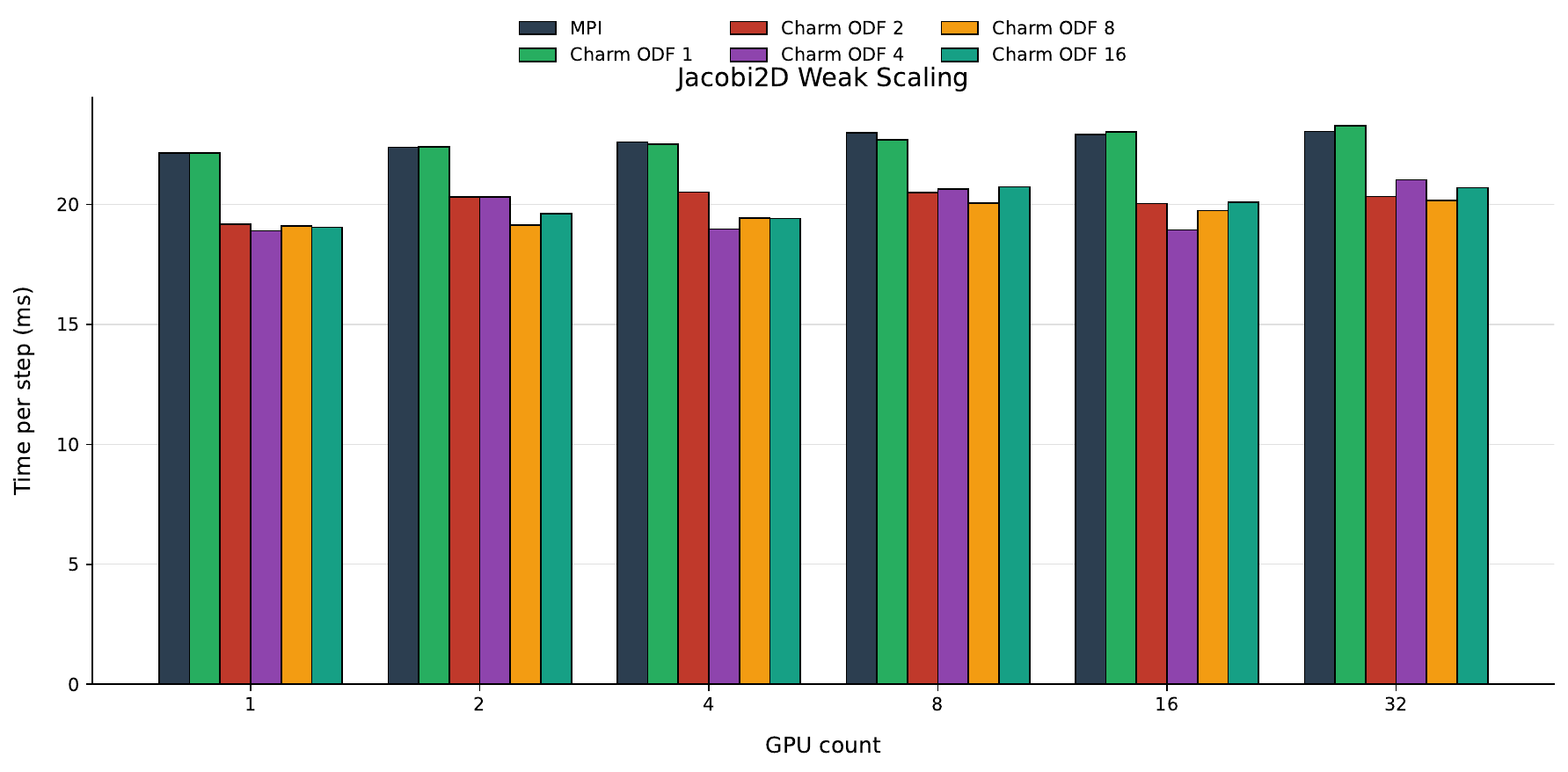}
    \caption{jacobi2d weak scaling on MI250X}
    \label{fig:jacobi2d_weak_scaling_amd}
\end{figure}

\begin{figure}
    \centering
    \includegraphics[width=1\linewidth]{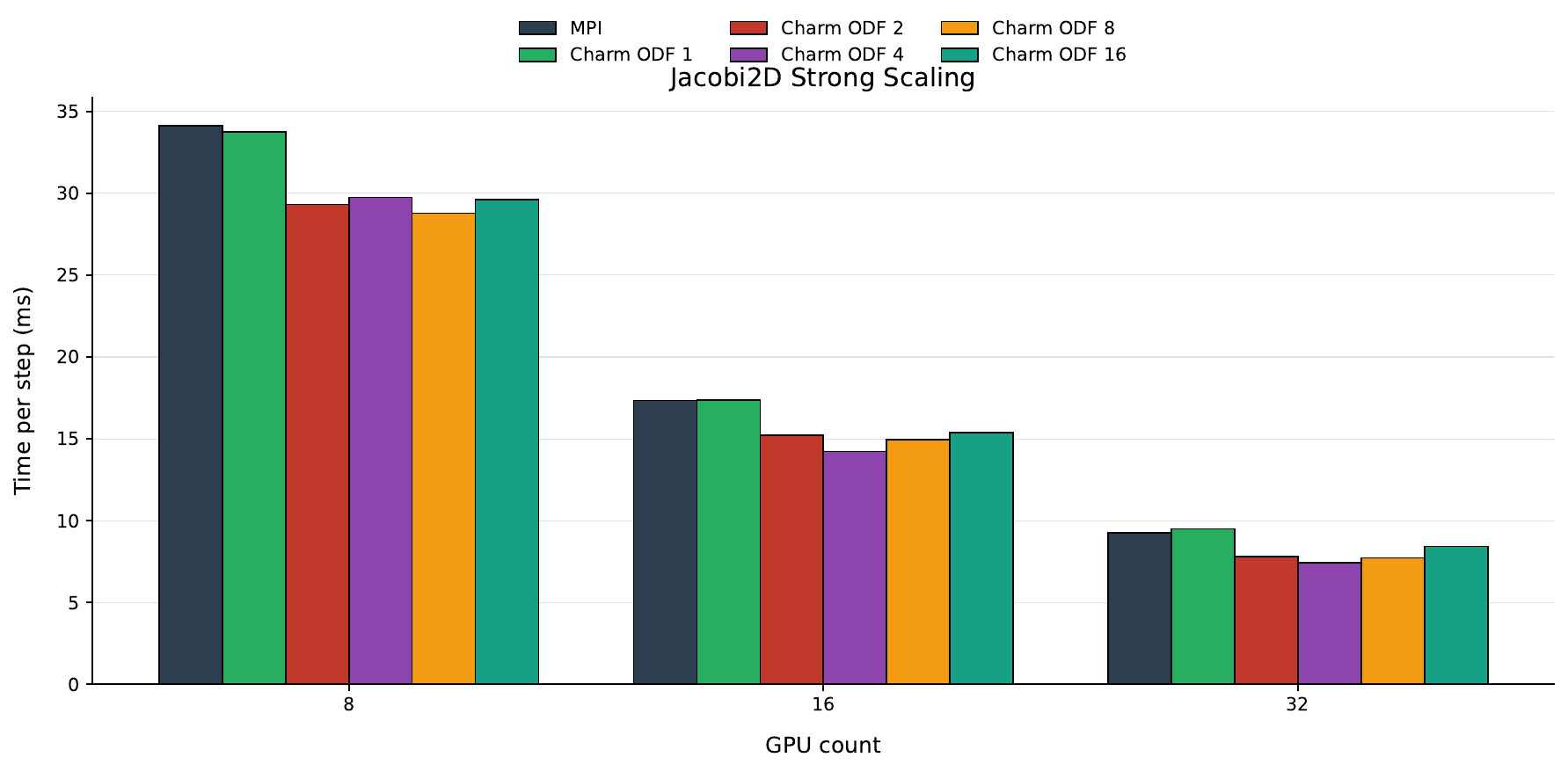}
    \caption{jacobi2d strong scaling on MI250X}
    \label{fig:jacobi2d_strong_scaling_amd}
\end{figure}


\subsubsection{MiniMD} 
MiniMD is a molecular dynamics proxy application that follows many of the same design principles as the parallel MD code LAMMPS. MiniMD uses spatial decomposition MD, where each Charm++ chare or MPI rank arranged in a grid owns a subset of the simulation domain. The application also exhibits a mixture of communication patterns due to its different communication routines. We run MiniMD for 100 iterations, disabling reverse communicate and safe-exchange checking.

\begin{figure}
    \centering
    \includegraphics[width=1\linewidth]{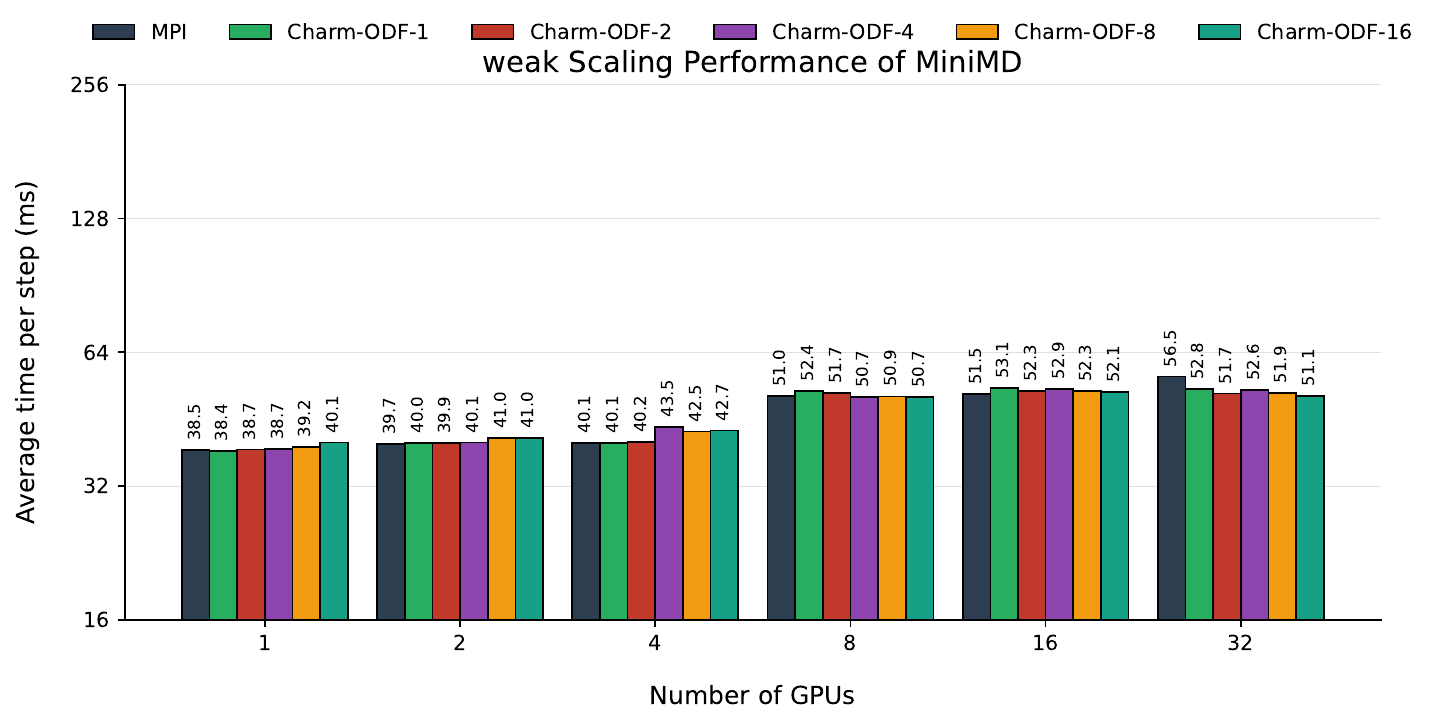}
    \caption{miniMD weak scaling on A40}
    \label{fig:minimd_weak_scaling}
\end{figure}

\textbf{Weak scaling on Nvidia.} We start with a base domain size of \(100 \times 100 \times 200\) and alternately scale each dimension at every scaling step, scaling from 1 GPU to 32 GPUs across 4 nodes. The overall trend for 1-16 GPUs is similar to that observed for Jacobi2D, with runtimes remaining comparable across MPI and Charm++ implementations with different ODF configurations. However, as shown in Figure~\ref{fig:minimd_weak_scaling}, Charm++ begins to outperform MPI at larger scales. At 32 GPUs, higher ODF configurations achieve a \(6.5\%-9.5\%\) improvement in runtime compared to MPI, suggesting that overdecomposition provides benefits through improved communication--computation overlap. The performance difference between ODF-1 and MPI, however, still requires further analysis.

\textbf{Strong scaling on NVIDIA.} We perform strong scaling experiments with a fixed domain size of \(300 \times 300 \times 300\), scaling from 4 GPUs (1 node) to 64 GPUs (8 nodes). Similar to the weak scaling results, the MPI and Charm++ implementations with different ODF configurations exhibit broadly comparable performance across the scaling range. However, at 32 and 64 GPUs, the ODF-16 configuration begins to perform slightly worse than intermediate ODFs. This behavior is likely due to the much smaller per-GPU problem sizes at higher scales, where excessive overdecomposition introduces runtime overheads that are not sufficiently hidden.

\begin{figure}
    \centering
    \includegraphics[width=1\linewidth]{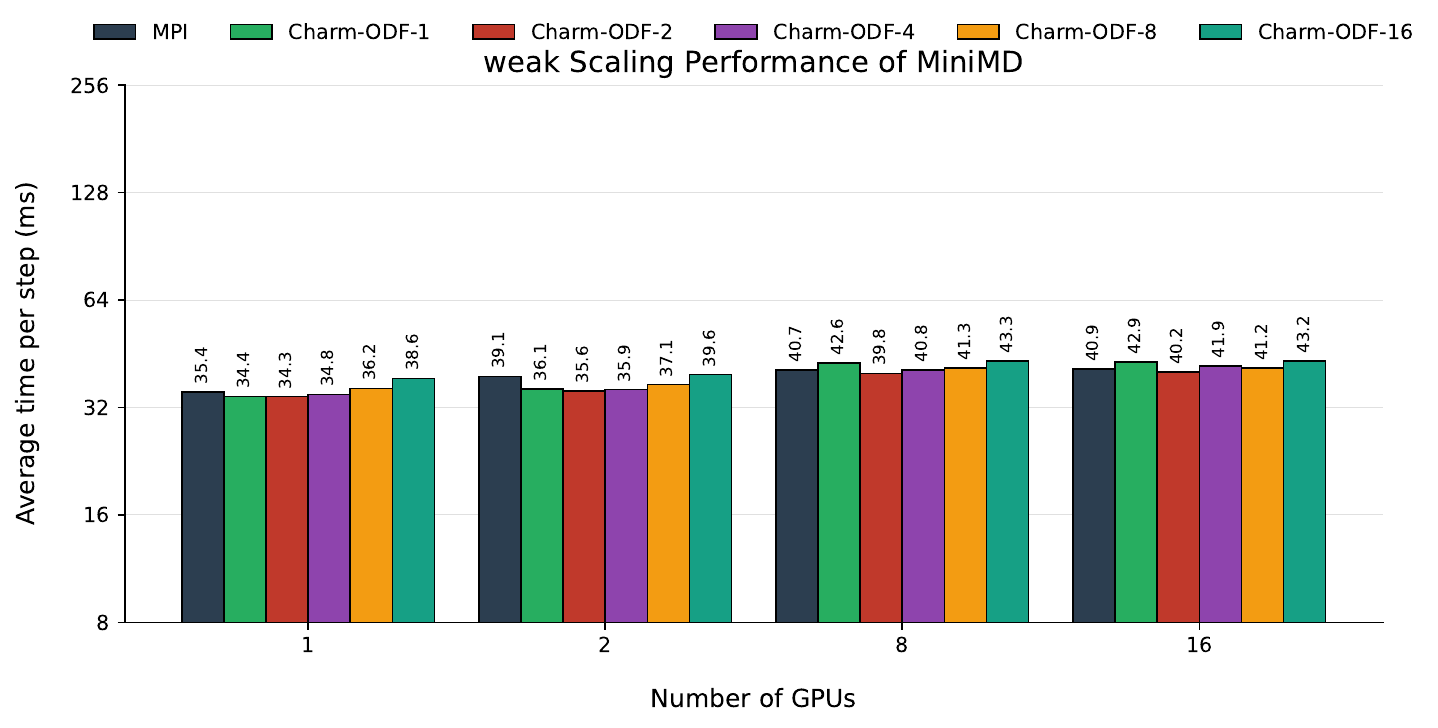}
    \caption{miniMD weak scaling on MI250X}
    \label{fig:minimd_weak_scaling_amd}
\end{figure}

\textbf{Weak scaling on AMD.} We use the same base domain size as in the NVIDIA experiments and scale MiniMD from 1 GPU to 16 GPUs (2 nodes). As shown in Figure~\ref{fig:minimd_weak_scaling_amd}, there is no noticeable performance gap between the Charm++ and MPI implementations, and the behavior is similar to what was observed for Nvidia.

\begin{figure}
    \centering
    \includegraphics[width=1\linewidth]{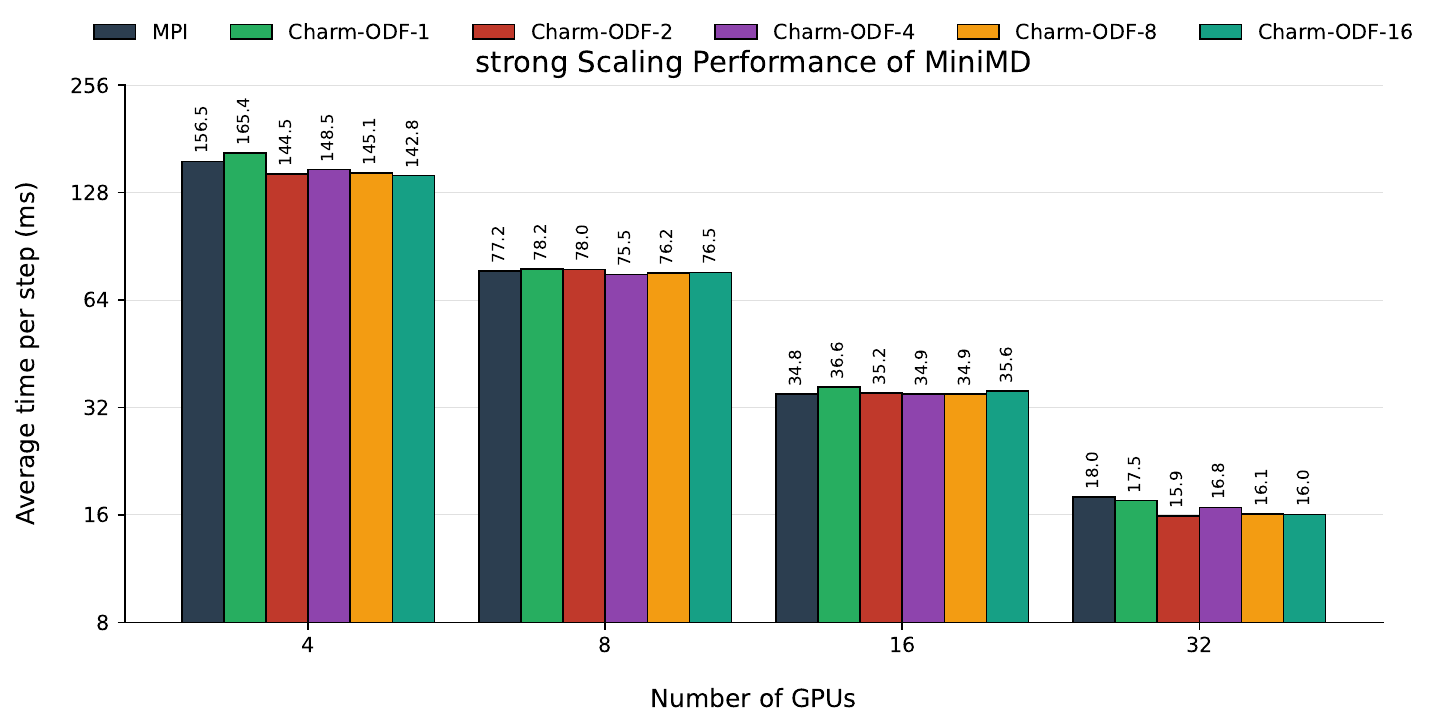}
    \caption{miniMD strong scaling on MI250X}
    \label{fig:minimd_strong_scaling_amd}
\end{figure}

\textbf{Strong scaling on AMD.} We perform strong scaling experiments with a fixed domain size same as the Nvidia runs, scaling from 1 GPU to 16 GPUs (2 nodes). The results in Figure~\ref{fig:minimd_strong_scaling_amd} indicate that Charm++ and MPI perform on par with each other, a trend that aligns with the Nvidia observations.

\begin{figure}
    \centering
    \includegraphics[width=1\linewidth]{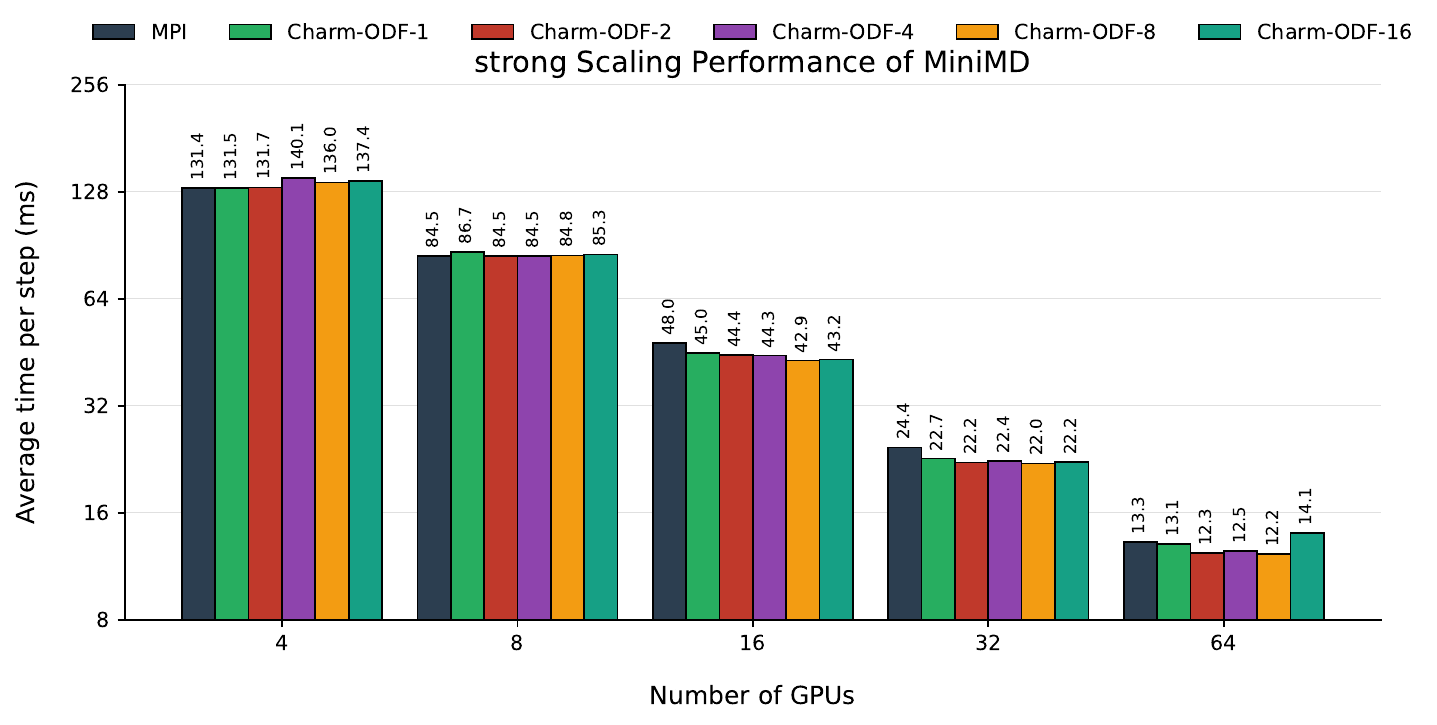}
    \caption{miniMD strong scaling on A40}
    \label{fig:minimd_strong_scaling}
\end{figure}

\subsubsection{LULESH}
LULESH (2.0) is a shock-hydrodynamics mini-application that solves the Sedov blast wave problem. It initializes a 3D hexahedral mesh of arbitrary size and partitions the domain across Charm++ chares or MPI ranks, with each rank responsible for a subdomain and for exchanging state across boundary elements. The application can also model load imbalance across domains, making it a useful benchmark for future load-balancing studies. The current MPI and Charm++ implementations are limited to decompositions with a perfect-cube number of domains.

Because of this perfect-cube constraint, the ODF often takes on fractional values. For example, running on 32 GPUs with a target ODF of 4 would ideally require 128 chares. Since 128 is not a perfect cube, the application instead uses the nearest perfect cube, $5^3 = 125$, resulting in an effective ODF of $125 / 32 \approx 3.91$. We run LULESH for 50 to 100 iterations depending on the scaling setup.

Additionally, the global mesh is constrained to have equal side lengths in all three dimensions, requiring each dimension to be scaled by \(\sqrt[3]{2}\) (\(\approx 1.25\)) whenever the compute is doubled during weak scaling. This restricts the MPI runs we can get to only perfect-cube number of GPUs. We still do charm++ runs for the entire range of scaling.

\textbf{Weak scaling on NVIDIA.} We evaluate weak scaling by increasing the grid dimensions as described earlier starting from a grid size of \(256\times256\times256\) to maintain a constant workload per GPU, scaling from 1 GPU to 32 GPUs across 8 nodes. The integral ODF configurations closely follow MPI performance across the scaling range, exhibiting trends similar to those observed for Jacobi2D and MiniMD. In contrast, the fractional ODF configurations consistently perform worse at most scales, indicating possible imbalance introduced by chare placement and mapping. This behavior warrants further investigation.

\textbf{Strong scaling on NVIDIA.} For strong scaling, we use a fixed domain size of \(480\times480\times480\) and scale from 4 GPUs (1 node) to 64 GPUs (16 nodes). At smaller scales, the Charm++ implementations continue to track MPI performance closely across different ODF configurations. However, at 32 and 64 GPUs, the higher integral ODFs begin to show a slight degradation in performance relative to intermediate ODFs, similar to the behavior observed in MiniMD. The reduced per-GPU problem size at these scales likely makes the overheads from excessive overdecomposition more pronounced. Fractional ODF configurations again remain consistently slower throughout the scaling range.

\begin{figure}
    \centering
    \includegraphics[width=1\linewidth]{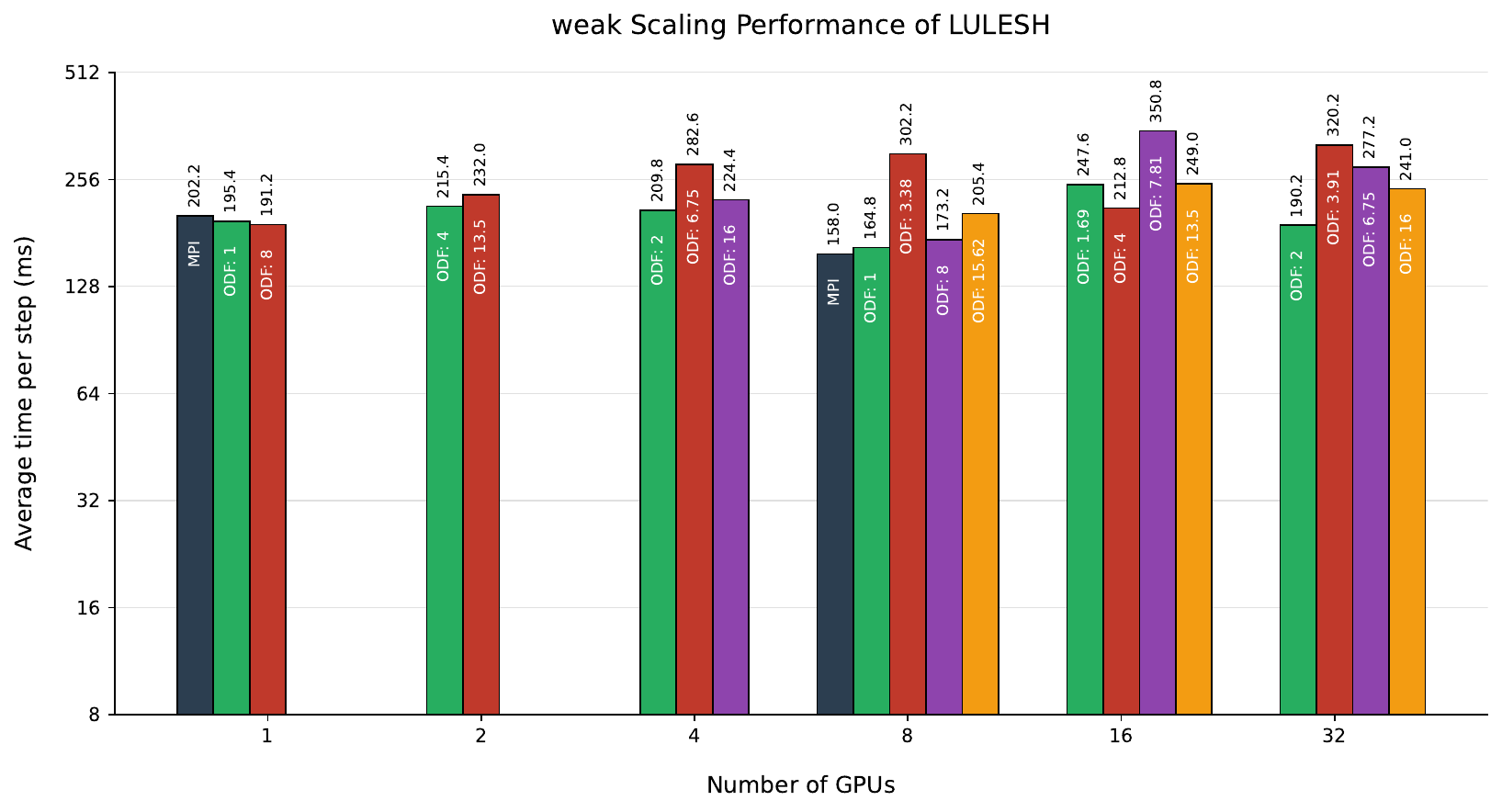}
    \caption{lulesh weak scaling on MI250X}
    \label{fig:lulesh_weak_scaling_amd}
\end{figure}

\textbf{Weak scaling on AMD.} On AMD systems, weak scaling experiments are performed from 1 GPU to 32 GPUs (4 nodes) with the base domain size of \(320\times320\times320\). Figure~\ref{fig:lulesh_weak_scaling_amd} shows almost similar runtimes between charm++ and MPI with some performance loss shown with higher ODFs. This behavior may be partially explained by limitations in the event-based IPC signaling mechanism used for intra-node HIP communication, which currently requires additional synchronization. 

\textbf{Strong scaling on AMD.} Strong scaling on AMD uses a fixed domain size same as the Nvidia runs, while scaling from 4 GPUs to 32 GPUs (4 nodes). The overall behavior mirrors the AMD weak scaling results, with Charm++ continuing to exhibit noticeable overheads relative to MPI for larger ODFs.

\begin{figure}
    \centering
    \includegraphics[width=1\linewidth]{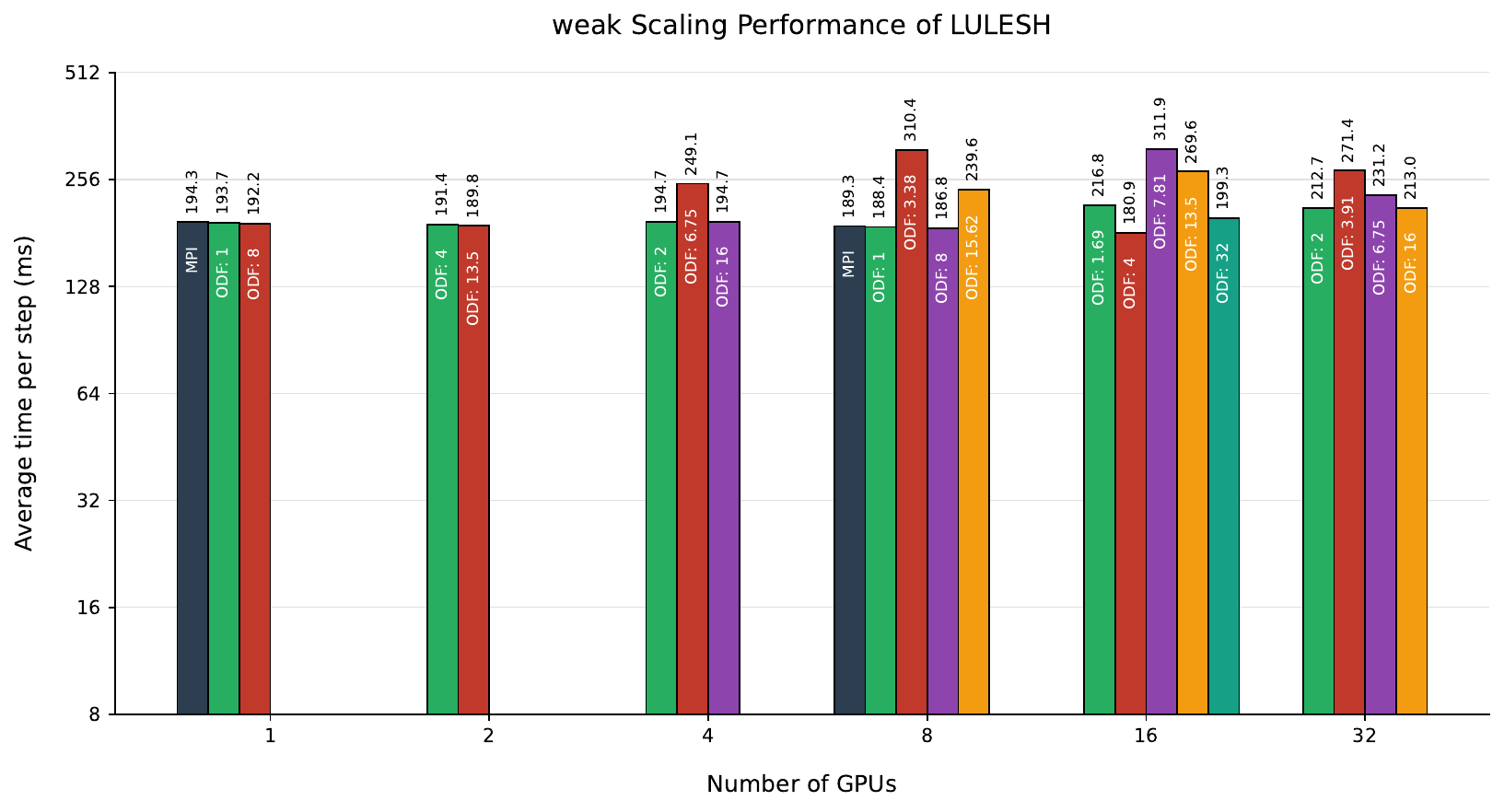}
    \caption{lulesh weak scaling on A40}
    \label{fig:lulesh_weak_scaling}
\end{figure}

\begin{figure}
    \centering
    \includegraphics[width=1\linewidth]{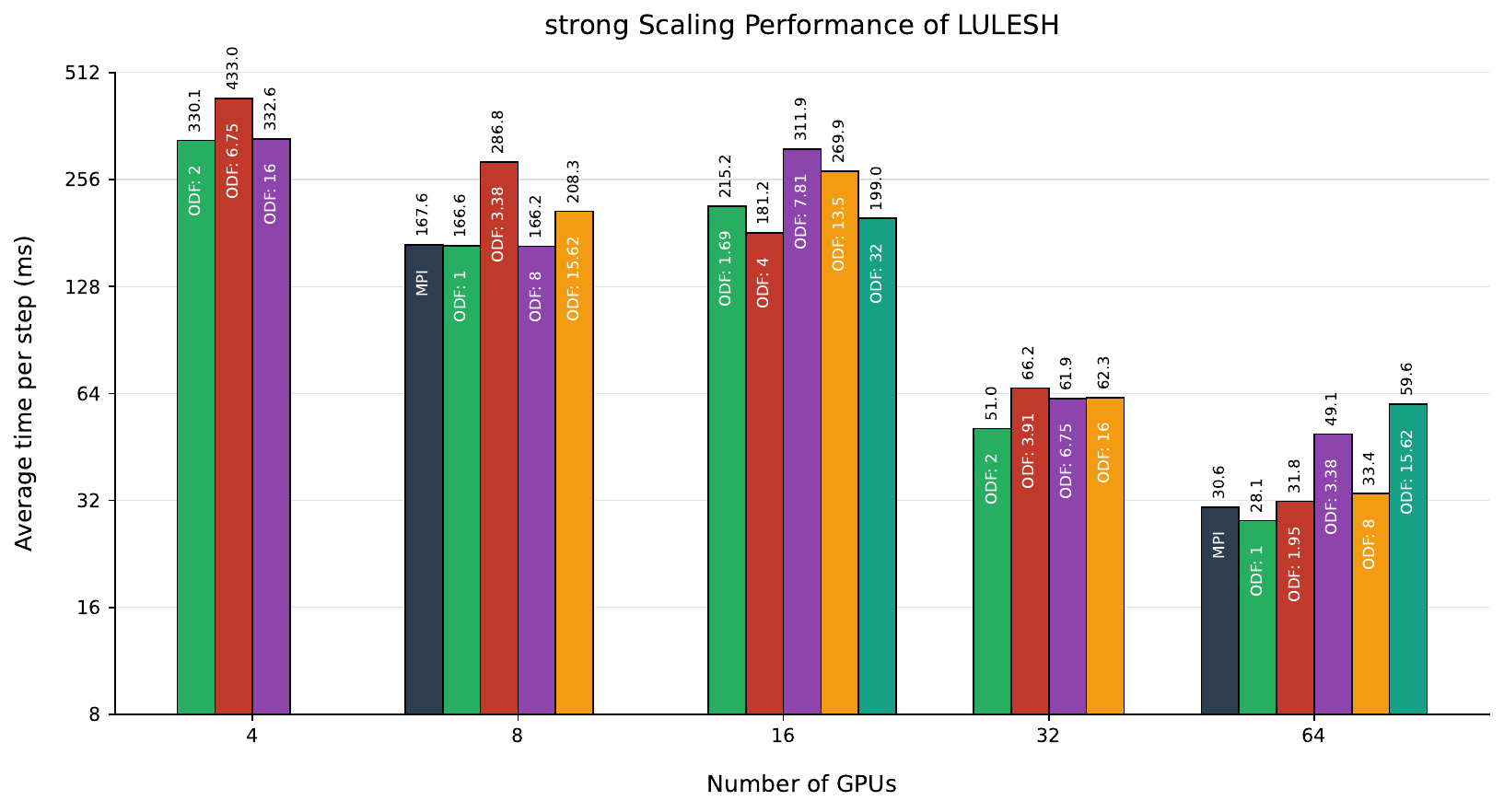}
    \caption{lulesh strong scaling on A40}
    \label{fig:lulesh_strong_scaling}
\end{figure}

\begin{figure}
    \centering
    \includegraphics[width=1\linewidth]{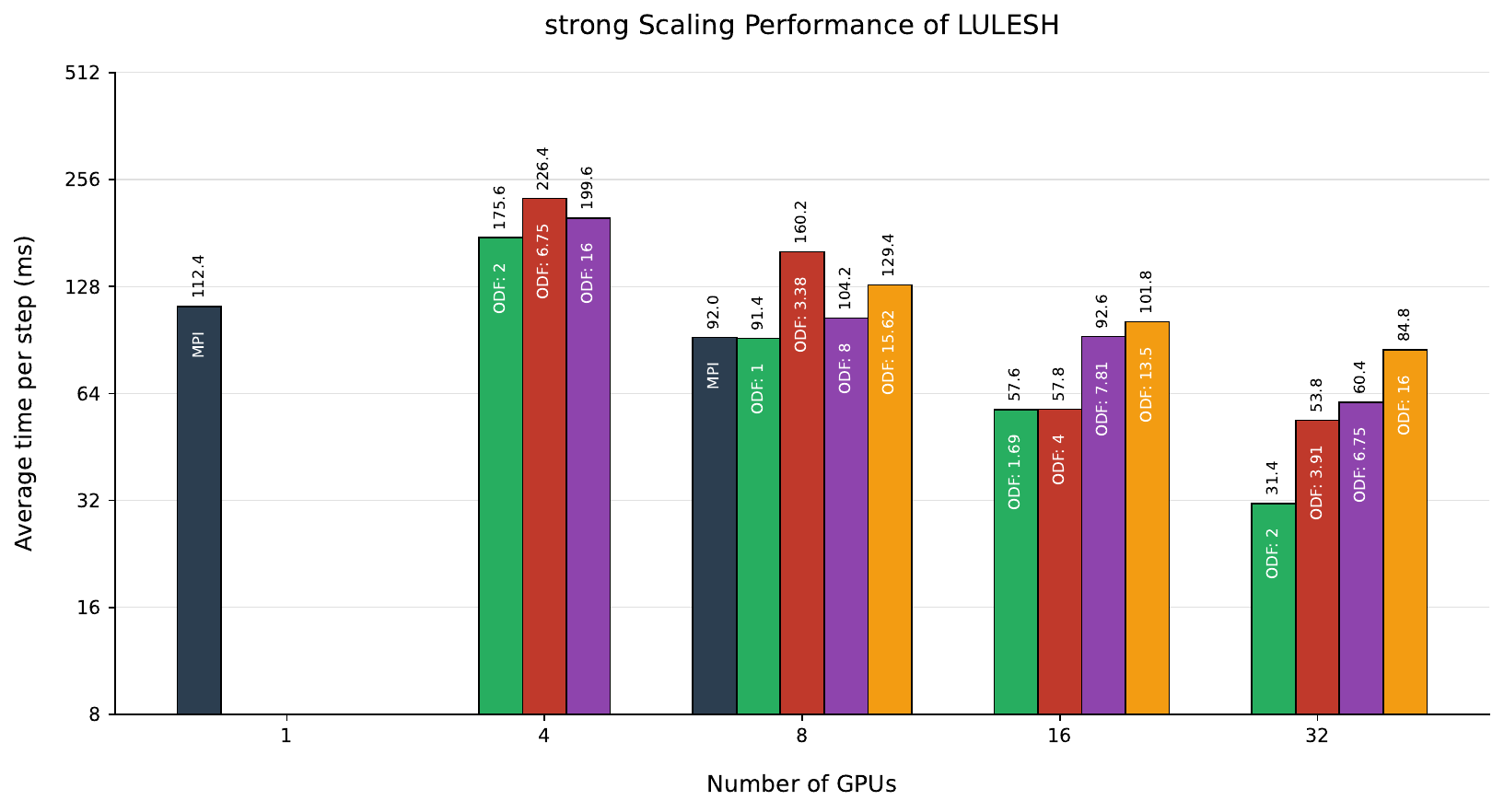}
    \caption{lulesh strong scaling on MI250X}
    \label{fig:lulesh_strong_scaling_amd}
\end{figure}

%% file: sections/conclusion.tex
\section{Discussion and Future Work}

We explored the overhead of overdecomposition in its computational and communication aspects. For computation, overdecomposition leads to smaller computation per kernel call. However, by employing efficient asynchronous techniques for kernel completion and continuations, utilizing multiple cores and chares to pipeline and overlap launch overhead, we demonstrated it is possible to run overdecomposed kernels at very low grainsizes.

Communication-wise, for overdecomposed programs to run efficiently, it is essential to efficiently handle multiple cases of source-destination pairs without requiring moving data via host. We implemented device-to-device data transfer in all code-paths in Charm++, based on the runtime's knowledge of the destination location, which normally is pushed to the lowest level of communication layer, where the layer, such as UCX may use its internal mechanism to effect the data transfer often with copying. We demonstrated that pipelined communication, and data driven scheduling allows us to support overdecomposition effectively. 

The three mini - applications demonstrated that overdecomposed applications run with about the same performance as non-overdecomposed ones, including MPI-only as well. 

There is still more research and evaluation needed to make progress along this direction. First, we need to study highly fine-grained benchmarks, such as biomolecular simulations, and astronomy codes, such as n-body gravity codes. These may require additional mechanisms to mitigate overheads.

More importantly, all of this only establishes the {\em potential} of overdecomposition. To realize and demonstrate its utility, specifically for dynamic load balancing and shrink/expand in the cloud context,  it is necessary to support chare migration efficiently, and to do accurate load estimation, each with its own challenges. 

The contributions of this paper create a foundation on which these future capabilities can be built. 

\section{Acknowledgments}

This work used Delta at NCSA through allocation ASC-050025 from the Advanced Cyberinfrastructure Coordination Ecosystem: Services \& Support (ACCESS) program, which is supported by U.S. National Science Foundation grants \#2138259, \#2138286, \#2138307, \#2137603, and \#2138296.

This research used the Delta advanced computing and data resource which is supported by the National Science Foundation (award OAC 2005572) and the State of Illinois. Delta is a joint effort of the University of Illinois Urbana-Champaign and its National Center for Supercomputing Applications.

This research used resources of the Oak Ridge Leadership Computing Facility at the Oak Ridge National Laboratory, which is supported by the Advanced Scientific Computing Research programs in the Office of Science of the U.S. Department of Energy under Contract No. DE-AC05-00OR22725.

%% file: group.bib
@article{NamdJCC05,
  author={James C. Phillips and Rosemary Braun and Wei Wang and James Gumbart and Emad Tajkhorshid and Elizabeth Villa and Christophe Chipot and Robert D. Skeel and Laxmikant Kal\'e and Klaus Schulten},
  title={{S}calable molecular dynamics with {NAMD}},
  journal={Journal of Computational Chemistry},
  volume={26},
  number={16},
  pages={1781-1802},
  year=2005,
  publisher={2005 Wiley Periodicals, Inc.},
}

@inproceedings{2007_ChaNGaScaling,
  author = "Pritish Jetley and Filippo Gioachin and Celso Mendes and Laxmikant V. Kale and Thomas R. Quinn",
  title = "{Massively parallel cosmological simulations with ChaNGa}",
  booktitle = "Proceedings of IEEE International Parallel and Distributed Processing Symposium 2008",
  year = "2008",
}

@inproceedings{sc14charm,
  title="{Parallel Programming with Migratable Objects: Charm++ in Practice}",
  author={Acun, Bilge and Gupta, Abhishek and Jain, Nikhil and Langer, Akhil and Menon, Harshitha and Mikida, Eric and Ni, Xiang and Robson, Michael and Sun, Yanhua and Totoni, Ehsan and Wesolowski, Lukasz and Kale, Laxmikant},
 series = {SC},
 year = {2014},
}

@inproceedings{CharmppOOPSLA93,
  author        = "{Kal\'{e}}, L.V. and Krishnan, S.",
  title         = "{CHARM++: A Portable Concurrent Object Oriented System
                   Based on C++}",
  editor        = "Paepcke, A.",
  fulleditor    = "Paepcke, Andreas",
  pages         = "91--108",
  Month         = "September",
  Year          = "1993",
  booktitle     = "{Proceedings of OOPSLA'93}",
  publisher     = "{ACM Press}",
}


%% file: references.bib
@inproceedings{LCI,
author = {Yan, Jiakun and Snir, Marc},
title = {LCI: a Lightweight Communication Interface for Efficient Asynchronous Multithreaded Communication},
year = {2025},
isbn = {9798400714665},
publisher = {Association for Computing Machinery},
address = {New York, NY, USA},
url = {https://doi.org/10.1145/3712285.3759881},
doi = {10.1145/3712285.3759881},
booktitle = {Proceedings of the International Conference for High Performance Computing, Networking, Storage and Analysis},
pages = {1043–1059},
numpages = {17},
location = {
},
series = {SC '25}
}

@INPROCEEDINGS{Choi22,
  author={Choi, Jaemin and Richards, David F. and Kale, Laxmikant V.},
  booktitle={2022 IEEE International Parallel and Distributed Processing Symposium Workshops (IPDPSW)}, 
  title={Improving Scalability with GPU-Aware Asynchronous Tasks}, 
  year={2022},
  volume={},
  number={},
  pages={569-578},
  doi={10.1109/IPDPSW55747.2022.00097}}

@article{KokkosCore2014,
  title = "Kokkos: Enabling manycore performance portability through polymorphic memory access patterns ",
  journal = "Journal of Parallel and Distributed Computing ",
  volume = "74",
  number = "12",
  pages = "3202 - 3216",
  year = "2014",
  note = "Domain-Specific Languages and High-Level Frameworks for High-Performance Computing ",
  issn = "0743-7315",
  doi = "https://doi.org/10.1016/j.jpdc.2014.07.003",
  url = "http://www.sciencedirect.com/science/article/pii/S0743731514001257",
  author = "H. Carter Edwards and Christian R. Trott and Daniel Sunderland"
}

@article{HPX, doi = {10.21105/joss.02352}, url = {https://doi.org/10.21105/joss.02352}, year = {2020}, publisher = {The Open Journal}, volume = {5}, number = {53}, pages = {2352}, author = {Kaiser, Hartmut and Diehl, Patrick and Lemoine, Adrian S. and Lelbach, Bryce Adelstein and Amini, Parsa and Berge, Agustín and Biddiscombe, John and Brandt, Steven R. and Gupta, Nikunj and Heller, Thomas and Huck, Kevin and Khatami, Zahra and Kheirkhahan, Alireza and Reverdell, Auriane and Shirzad, Shahrzad and Simberg, Mikael and Wagle, Bibek and Wei, Weile and Zhang, Tianyi}, title = {HPX - The C++ Standard Library for Parallelism and Concurrency}, journal = {Journal of Open Source Software} }

@inproceedings{Legion,
author = {Bauer, Michael and Treichler, Sean and Slaughter, Elliott and Aiken, Alex},
title = {Legion: expressing locality and independence with logical regions},
year = {2012},
isbn = {9781467308045},
publisher = {IEEE Computer Society Press},
address = {Washington, DC, USA},
booktitle = {Proceedings of the International Conference on High Performance Computing, Networking, Storage and Analysis},
articleno = {66},
numpages = {11},
location = {Salt Lake City, Utah},
series = {SC '12}
}

@INPROCEEDINGS{raja,
  author={Beckingsale, David A. and Burmark, Jason and Hornung, Rich and Jones, Holger and Killian, William and Kunen, Adam J. and Pearce, Olga and Robinson, Peter and Ryujin, Brian S. and Scogland, Thomas RW},
  booktitle={2019 IEEE/ACM International Workshop on Performance, Portability and Productivity in HPC (P3HPC)}, 
  title={RAJA: Portable Performance for Large-Scale Scientific Applications}, 
  year={2019},
  volume={},
  number={},
  pages={71-81},
  doi={10.1109/P3HPC49587.2019.00012}}

@inproceedings{uintah,
author = {Holmen, John K. and Garc\'{\i}a, Marta and Sanderson, Allen and Bagusetty, Abhishek and Berzins, Martin},
title = {Lessons Learned and Scalability Achieved When Porting Uintah to DOE Exascale Systems},
year = {2024},
isbn = {978-3-031-90199-7},
publisher = {Springer-Verlag},
address = {Berlin, Heidelberg},
url = {https://doi.org/10.1007/978-3-031-90200-0_19},
doi = {10.1007/978-3-031-90200-0_19},
booktitle = {Euro-Par 2024: Parallel Processing Workshops: Euro-Par 2024 International Workshops, Madrid, Spain, August 26–30, 2024, Proceedings, Part I},
pages = {231–242},
numpages = {12},
location = {Madrid, Spain}
}
